\def\re#1{(\ref{#1})}
\def\beq{\begin{equation}}
\def\eeq{\end{equation}}
\def\beeq{\begin{eqnarray}}
\def\beeqn{\begin{eqnarray*}}
\def\eeeq{\end{eqnarray}}
\def\eeeqn{\end{eqnarray*}}
\def\de{\delta}                 \def\D{\Delta}

\def\l{\lambda}                 
\def\m{\mu}
\def\n{\nu}

\newcommand{\DD}{{\cal D}}

\newcommand{\OO}{{\cal O}}

\newcommand{\WW}{{\cal W}}

\newcommand{\lp}{\left(}
\newcommand{\rp}{\right)}
\renewcommand{\lq}{\left[}
\renewcommand{\rq}{\right]}

\newcommand{\no}{\nonumber}
\newcommand{\ph}{\phantom} 

\def\tr{\,\mbox{Tr}\,}
\def\frac#1#2{ {{#1} \over {#2} }}

\def\p{\partial}
\def\ie{\hbox{\it i.e.}{ }}

\newcommand{\unity}{1\kern-.65mm \mbox{\form l}}
\newcommand{\ks}{\mbox{\form l}\kern-.6mm \mbox{\form K}}
\newcommand{\A}{A \raise0.5mm\hbox{\kern-1.8mm /}}
\def\pmb#1{\leavevmode\setbox0=\hbox{$#1$}\kern-.025em\copy0\kern-\wd0
\kern-.05em\copy0\kern-\wd0\kern-.025em\raise.0433em\box0}
\def\D{\hbox{\hbox{${D}$}}\kern-1.9mm{\hbox{${/}$}}}
\def\kbar{\hbox{$k$}\kern-0.2true cm\hbox{$/$}}
\def\nbar{\hbox{$n$}\kern-0.23true cm\hbox{$/$}}
\def\pbar{\hbox{$p$}\kern-0.18true cm\hbox{$/$}}
\def\nhbar{\hbox{$\hat n$}\kern-0.23true cm\hbox{$/$}}

\documentstyle[epsfig,preprint,aps,floats,amssymb]{revtex}
\begin{document}
\draft
\newfont{\form}{cmss10}

\title{Perturbative Wilson loop in two-dimensional non-commutative
Yang-Mills theory}
\author{A. Bassetto$^1$, G. Nardelli$^2$ and A. Torrielli$^1$}
\address{$^1$Dipartimento di Fisica ``G.Galilei", Via Marzolo 8, 35131
Padova, Italy\\
INFN, Sezione di Padova, Italy \texttt{(bassetto, torrielli@pd.infn.it)}\\
$^2$Dipartimento di Fisica, Universit\`a di Trento,
38050 Povo (Trento), Italy \\ INFN, Gruppo Collegato di Trento, Italy 
\texttt{(nardelli@science.unitn.it)}}
\maketitle
\begin{abstract}
We perform a perturbative ${\cal O}(g^4)$ Wilson loop calculation for the $U(N)$
Yang-Mills theory defined on non-commutative one space - one time dimensions. 
We choose the light-cone gauge and compare the results obtained when using the 
Wu-Mandelstam-Leibbrandt ($WML$) and the Cauchy principal value ($PV$) prescription
for the vector propagator. In the $WML$ case the $\theta$-dependent term
is well-defined and
regular in the limit $\theta \to 0$, where the commutative theory is recovered;
it provides a non-trivial example of a consistent calculation when non-commutativity
involves the time variable. In the $PV$ case, unexpectedly, the result differs from 
the $WML$ one only by the addition of two singular terms with a trivial 
$\theta$-dependence. We find this feature intriguing, when remembering that, in
ordinary theories on compact manifolds, the difference between the two cases can be 
traced back to the contribution of topological excitations. 
\end{abstract}
\vskip 0.5truecm
DFPD 01/TH/29, UTF/445.



\vfill\eject

\narrowtext

\section{Introduction}
\noindent
There has been recently a growing interest in field theories defined on 
non-commutative spaces. 
Although their ultimate ``physical'' motivation is 
provided, in our opinion, by their tight relation
with  some limiting cases of string theories \cite{cds,dh,seiwi}, the 
very possibility 
of exploring some specific non-local field theories in a
systematic way in search of unexpected properties is fascinating on its own.
These field theories are indeed non-local and  non-locality has dramatic 
consequences on their 
basic dynamical features \cite{minwa,suski}.

A concrete way of turning ordinary theories into non-commutative ones is 
to replace  the 
usual multiplication of fields in the Lagrangian with the
$\star$-product. 
This product is constructed by means of a real antisymmetric matrix 
$\theta^{\mu\nu}$ which parameterizes  non-commutativity of Minkowski 
space-time:
\beq
\label{alge}
[x^\mu,x^\nu]=i\theta^{\mu\nu}\quad\quad\quad\quad \mu,\nu=0,..,D-1.
\eeq
The $\star$-product of two fields $\phi_1(x)$ and $\phi_2(x)$ is defined as
\beq
\label{star}
\phi_1(x)\star\phi_2(x)=\exp \lq \frac{i}2 \, \theta^{\mu\nu}\frac{\partial}
{\partial x_1^\mu}
\frac{\partial}{\partial x_2^\nu}\rq \phi_1(x_1)\phi_2(x_2)|_{x_1=x_2=x}
\eeq
and  leads to terms in the action with an infinite number of derivatives of 
fields which makes the theory non-local. Then one may wonder whether the theory 
would still comply with unitarity requirements.

Unitarity of scalar field theories, in the presence of non-commutativity, has 
been discussed in Ref.~\cite{gomis}: the authors 
explicitly check that Cutkoski's rules are correct when $\theta^{\mu\nu}$ 
is of the ``spatial'' type, \ie $\theta^{0i}=0$. This  exactly
corresponds to the case in 
which an elegant embedding into string theory is possible: low-energy 
excitations of a $D$-brane in a magnetic background are in fact described by 
field theories with space non-commutativity \cite{seiwi},
\cite{stringhi}.

On the other hand theories with $\theta^{0i}\neq 0$ have an infinite number
of time derivatives and are non-local in time: in this situation
unitarity may be in jeopardy, the concept of
Hamiltonian losing, in some sense, its meaning (see however \cite{kamimura}).

Again, this fact is not surprising when observed
from the string theory point of view: $\theta^{0i}\neq 0$ is obtained in
the presence of an electric background. The truncation of such a string theory
to its massless sector is not consistent in this case 
 \cite{om} (see also \cite{barb}).
The breakdown of unitarity in time-like scalar non-commutative theories has been
considered in Ref.~\cite{alva}.

In the context of non-commutative quantum Yang-Mills theories, 
unitarity has been recently discussed in Ref.\cite{bgnv}.

In ordinary theories, a typical probe to check unitarity is provided 
by the exponentiation of a Wilson loop. A perturbative computation has 
been widely used to check  unitarity, assuming 
gauge invariance, or viceversa \cite{test}.
 
To extend this test to non-commutative theories is highly problematic, 
even in the spatial case. As a matter of fact the definition of the loop
via a non-commutative path-ordering \cite{gross,alvar,dorn}, 
has so far received a physical interpretation in the presence of
matter fields as a wave function of composite operators only in a
lattice formulation \cite{ambj}. 

Exponentiation itself has not been proven to our knowledge,
even in the spatial case. Nevertheless
in Ref.(\cite{bgnv}) a perturbative ${\cal O}(g^4)$
generalization to the spatial non-commutative case of the familiar results
in the usual theory has shown that exponentiation persists
in spite of the phases which reflect non-commutativity.

A particularly interesting case in ordinary $U(N)$ gauge theories occurs in
one-space, one-time dimensions ($YM_{1+1}$): thanks to the invariance
under area-preserving diffeomorphisms, the Wilson loop can be exactly
computed \cite{stau}. 

If the calculation of the loop is performed at all orders in perturbation theory using 
the light-cone gauge $A_{-}=0$, two
{\it different} functions of the area are obtained, according to the two different 
prescriptions chosen for the vector propagator in momentum space, namely
\beq
\label{hooft}
D_{++}=i\ [k_{-}^{-2}]_{PV}
\eeq
and
\beq
\label{mand}
D_{++}=i\ [k_{-}+i\epsilon k_{+}]^{-2},
\eeq
$PV$ denoting the Cauchy principal value.
The two expressions above are usually referred in the literature as 't Hooft \cite{hoo}
and Wu-Mandelstam-Leibbrandt ($WML$) \cite{wum} propagators. They correspond to 
two different ways of quantizing the theory, namely by means of a light-front
or of an equal-time algebra \cite{bbg}, respectively.

The loop for $U(N)$, computed according to 't Hooft, coincides 
with the exact one obtained 
on the basis of
geometrical considerations \cite{stau}
\beq
\label{looft}
{\cal W}= \exp (\frac{i}{2}g^2 N {\cal A}),
\eeq
and exhibits the Abelian-like exponentiation one expects on the basis of unitarity
arguments,
whereas the $WML$ propagator leads to a different, genuinely perturbative
result in which topological effects are disregarded \cite{anlu}
\beq
\label{wml}
{\cal W}_{WML}= \exp (\frac{i}{2}g^2 {\cal A}) L^{(1)}_{N-1}(-i g^2 {\cal A}),  
\eeq
$L^{(1)}_{N-1}$ being a Laguerre polynomial.

The $WML$ propagator can be Wick-rotated, thereby
allowing for an Euclidean treatment. This is indeed the normal procedure followed in
order to obtain the solution above. The continuation of the propagator
is instead impossible when using the $PV$ prescription.

One should however bear in mind that in the Euclidean formulation the original
Minkowski contour of integration $\Gamma$ is changed to a (complex) contour
$\Gamma^*$, according to the rule $(x_0,x_1) \to (ix_2,x_1)$. The area ${\cal A}$
is thereby converted into $i A$. From what we have said, the Euclidean formulation 
cannot be defined if we choose
the 't Hooft propagator.

\smallskip

One can now inquire to what extent these considerations can be generalized to a
non-commutative $U(N)$ gauge theory, always remaining in 1+1 dimensions.
This is just the subject of the present work.

In 1+1 dimensions, non-commutativity necessarily affects the time variable
leading to a potentially pathological theory.

As the Euclidean formulation has better regularity properties, we shall first 
consider the Euclidean non-commutative $YM$. In so doing however, the non-commutative
parameter has to undergo a transition to imaginary values as well, so that
the basic (deformed) algebra (\ref{alge}) is preserved \cite{gomis}. 

In the next Section the Euclidean Wilson loop will be computed 
at ${\cal O}(g^4)$. We shall find that it exhibits all the expected features,
namely pure area dependence and continuity in the limit of vanishing non-commutative
parameter. The limiting case of large non-commutative parameter 
(maximal non-commutativity)
will also be discussed as well as the problems one encounters when trying to go
beyond the ${\cal O}(g^4)$-result.

In Sect.3 we shall face the difficulties inherent to a Minkowski formulation.
We find, by performing a direct calculation in Minkowski space with the $WML$
propagator, that the Wilson loop is in agreement with the
Euclidean one, after both the area ${\cal A}$ and the parameter $\theta$
have undergone simultaneous analytic reflection ${\cal A}\to i{ A}$,
$\theta\to i\theta.$ 

One obtains in this way a regular result, even if the theory has
space-time non-commutativity. A similar behaviour occurs also in perturbative Green
functions, which keep their functional dependence on momenta,
in spite of different analyticity properties coming from the analytic
continuation to Minkowski variables \cite{gomis}.

An Euclidean formulation is instead impossible if the 't Hooft propagator
is considered, as no Wick rotation is allowed for it. An
${\cal O}(g^4)$ calculation  
performed directly for the Minkowski contour
produces singular integrals. Those integrals can be regularized by introducing a
cutoff, but the final expression will irremediably diverge when the cutoff is removed.

Final comments and possible future developments are discussed in Sect.4.

\section{The Euclidean Wilson loop}

\noindent
In this section we analyze the $U(N)$ Yang-Mills theory on a two-dimensional
non-commutative space. 
The classical Minkowski action reads 
\beq 
\label{action}
S=-\ \frac1{2} \int d^2x\,Tr \Big( F_{\m\n} \star  F^{\m\n} \Big)
\eeq
where the field strength $F_{\m\n}$ is given by
\beq
F_{\m\n}=\p_\m A_\n -\p_\n A_\m -ig (A_\m\star A_\n - A_\n\star A_\m)
\eeq
and $A_\m$ is a $N\times N$ hermitian matrix.
The $\star$-product was defined in  Eq.~\re{star}.
The action in Eq.~\re{action} is invariant under $U(N)$
non-commutative gauge transformations 
\beq
\label{gauge}
\de_\l A_\m= \p_\m \l -ig (A_\m\star\l -\l\star A_\m) \,.
\eeq
We quantize the theory in the light-cone gauge $n^{\mu}A_{\mu}\equiv A_{-}=0$,
the vector $n_{\mu}$ being light-like, $n^{\mu}\equiv\frac{1}{\sqrt 2}(1,-1).$
The gauge condition can be imposed either adding to the action the gauge-fixing term
\beq
S_{g.f.}=\frac1{\alpha} \int d^2x\,Tr\Big( (A_{-} \star  A_{-})(x) \Big)=
\frac1{\alpha} \int d^2x\,Tr \Big( (A_{-}(x))^2 \Big), \qquad \alpha \to 0,
\eeq
or by means of a Lagrange multiplier. 

It is known that
Faddeev-Popov ghosts decouple even in non-commutative theories \cite{sheikh},
whereas the field tensor has only one component $F_{-+}=\p_{-} A_{+}.$

The free vector propagator coincides with the one of the ordinary theory,
namely it has the expressions of Eqs.(\ref{hooft}) or (\ref{mand}),
according to the algebra used for quantizing the theory.

In two dimensions, when quantized in axial gauge, the theory looks indeed free; 
in particular the action
remains trivially invariant under area preserving diffeomorphisms; yet it can give
rise to non-trivial correlations by means of open Wilson lines and
to non trivial potentials between external sources starting from closed Wilson loops.

In the non-commutative case the  Wilson loop can be defined by means of the
Moyal product as \cite{gross,alvar}  
\beq
\label{wloop}
\WW[C]=\frac{1}{N}\int \DD A \, e^{iS[A]} \int d^2x\,\tr P_{\star} \exp \lp
ig \int_C A_{+} 
(x+\xi(s))\, d\xi^{+}(s)\rp \,,
\eeq
where $C$ is a closed contour in non-commutative space-time
parameterized by $\xi(s)$, with $0 \leq s \leq 1$, and $P_\star$
denotes non-commutative path ordering along $x(s)$ from left to right
with respect to increasing $s$ of $\star$-products of functions.
Gauge invariance requires integration over coordinates, which is
trivially realized when considering  vacuum averages \cite{dorn}.
 
The perturbative expansion of $\WW [C]$,
expressed by Eq.~\re{wloop}, reads 
\beeq \label{loopert}
\WW   [C]&=&\frac{1}{N}
\sum_{n=0}^\infty (ig)^n \int_{0}^1 ds_1 \ldots
\int_{s_{n-1}}^{1} ds_{n} 
\, \dot{x}_{-}(s_1)\ldots\dot{x}_{-}(s_{n})\no\\
&&\ph{
\sum_{n=0}^\infty (ig)^n } 
\langle 0\left|\tr{\cal T}\lq  A_{+}(x(s_1))
\star\ldots\star    A_{+}(x(s_{n}))\rq \right|0\rangle, 
\eeeq
and it is easily shown to be an even power series in $g$, so that we can write
\beq
\label{wpert}
\WW   [C]= 1 +g^2 \WW_2+g^4\WW_4+\OO(g^6)\,.
\eeq
Through an explicit evaluation one is convinced that the function
$\WW_2$ in Eq.~\re{wpert} is reproduced by the single-exchange diagram, 
which is exactly as in the ordinary $U(N)$ theory.

Therefore the first interesting quantity is $\WW_4$. It is obtained by
the exchange of two propagators which can either cross or uncross. In
the pairings $(12)(34)$ and $(23)(41)$ they do not cross, whereas
in the pairing $(13)(24)$ they do. Let us denote by $\WW_4^{(1)}$,
$\WW_4^{(2)}$ and $\WW_4^{(cr)}$ the contributions in the three cases.

It is not difficult to realize that $\WW_4^{(1)}$ and
$\WW_4^{(2)}$ are the same as the corresponding ones in the ordinary 
theory, being unaffected by the Moyal phase. We therefore concentrate 
our attention on $\WW_4^{(cr)}$ and consider for the time being an Euclidean
formulation.
By exploiting the invariance of $\WW_4$ under area-preserving diffeomorphisms,
which holds also in this non-commutative context, we consider the simple 
choice of a circular contour \cite{stau} 
\beq
\label{cerchio}
x(s)\equiv(x_1(s),x_2(s))=r(\cos(2\pi s),\sin(2\pi s)).
\eeq

Were it not for the presence of the Moyal phase, a tremendous simplification
would occur between the factor in the measure $\dot{x}_{-}(s)\dot{x}_{-}(s')$
and the basic correlator $<A_{+}(s)A_{+}(s')>$ \cite{stau}. The Moyal phase can be
handled in an easier way if we perform a Fourier transform, namely if
we work in the momentum space. The momenta are chosen to be Euclidean
and the non-commutative parameter imaginary $\theta\to i \theta$. 
In this way all the phase factors do not change their behaviour.

We start from Eq.(\ref{loopert}) and single out its ${\cal O}(g^4)$ contribution; 
we use WML propagators in the Euclidean form $(k_1-ik_2)^{-2}$ and parameterize
the vectors introducing polar variables. Then we are led to the expression
\begin{eqnarray}  
\label{crociato}
\WW_4^{(cr)} &=& \, r^4 \, \int_0^{1} d{s_1} \int_{s_1}^{1} d{s_2} \int_{s_2}^{1} 
d{s_3 }  \int_{s_3}^{1} d{s_4} \times \nonumber \\
&&\int_0^{\infty } \, {{dp}\over {p}} {{dq}\over {q}} 
\, \int_0^{2 \pi } d\psi \, d\chi \, \exp({- 2 i (\psi + \chi )}) 
\exp({2 i p \sin \psi \sin \pi (s_1 - s_3 )})\times \nonumber  \\
&& \exp({2 i q \sin \chi \sin \pi (s_2 - s_4 )})  \exp({i {{\theta }\over {r^2}} 
p \ q\ \sin [\psi - \chi + \pi (s_2 + s_4 -  s_1 - s_3)] })\nonumber \\
&=& { A}^2 \ F(\frac{\theta}{{ A}}).
\end{eqnarray}

Integrating over $\psi$ and $p$, we get
\begin{equation}  
\label{integrato}
\WW_4^{(cr)} \, = \, \pi r^4 \int [ds] \, \int_0^{\infty } {{dq} \over {q}} \, 
\ointop_{|z|=1}\, {{dz} \over {i {{z}^3}}} \, \, e^{-\ q  \sin [\pi (s_4 - s_2 )] (z - {{1} \over {z}})} \, {{1 - {{\gamma } \over {z}} e^{- i \pi \sigma }} \over {1 - \gamma z e^{i \pi \sigma }}},
\end{equation}
where  $\sigma = s_1 + s_3 - s_2 - s_4 $ and
$$ \gamma = \frac {\theta q }{2 r^2 \sin \pi (s_3 - s_1 )},\qquad\qquad
\int [ds] = \int_0^1 ds_1 \int_{s_1 }^1 ds_2 \int_{s_2 }^1 ds_3 \int_{s_3 }^1 ds_4 .$$ 

It is an easy calculation to check that the function $F$ is continuous at 
$\theta=0$ with $F(0)=\frac{1}{24}$, exactly corresponding to the value
of the commutative case obtained with the WML propagator
(see the ${\cal O}(g^4)$ term of Eq.(\ref{wml})). 

One can also evaluate the function $F$
in the large-$\theta$ limit. This is most easily done by considering the splitting
\begin{eqnarray}
\label{splitted}
\WW_4^{(cr)} \, &=& \, \pi r^4 \int [ds] \, \int_0^{\infty } {{dq} \over {q}} \, 
\ointop_{|z|=1}\, {{dz} \over {i {{z}^3}}} \, \, \Big( e^{-\ q  \sin [\pi (s_4 - s_2 )] (z - {{1} \over {z}})} - 1 \Big) \, {{1 - {{\gamma } \over {z}} e^{- i \pi \sigma }} \over {1 - \gamma z e^{i \pi \sigma }}}\nonumber \\
&+& \, \pi r^4 \int [ds] \, \int_0^{\infty } {{dq} \over {q}} \, \ointop_{|z|=1} \, {{dz} \over {i {{z}^3}}} \, \,  {{1 - {{\gamma } \over {z}} e^{- i \pi \sigma }} \over {1 - \gamma z e^{i \pi \sigma }}} .
\end{eqnarray}
In the first integral we are entitled to perform the large-$\theta$ limit, thereby
obtaining
\begin{eqnarray}
\label{appro}
{\cal I}_1 &\simeq& \pi r^4 \int [ds] \, \int_0^{\infty } {{dq} \over {q}} \, 
\ointop_{|z|=1}\, {{dz} \over {i {{z}^3}}} \, \, \Big( e^{-\ q  \sin [\pi (s_4 - s_2 )] (z - {{1} \over {z}})} - 1 \Big) \, {{e^{- 2 i \pi \sigma }} \over {z^2 }}\nonumber \\
&=&{{r^4 {\pi }^2 } \over {2}} \int [ds] e^{-2 \pi i (s_1 + s_3 - s_2 - s_4 )} \, = \, - {{r^4 } \over {16}} (1 - {{3 i} \over {\pi }} ).
\end{eqnarray}
The second integral in turn gives
\begin{eqnarray}
\label{apro}
{\cal I}_2 &=&  \pi r^4 \int [ds] \, \int_0^{\infty } {{dq} \over {q}} \, 
\ointop_{|z|=1} \, {{dz} \over {i {{z}^3}}} \, \,  {{1 - {{\gamma } \over {z}} e^{- i \pi \sigma }} \over {1 - \gamma z e^{i \pi \sigma }}} \nonumber \\
&=&
2\pi^2 r^4 \int [ds] \, \int_0^1\gamma d\gamma e^{2 i \pi \sigma }\, (1-{\gamma }^2)
\nonumber \\
&=& {{ r^4 {\pi }^2 } \over {2}} \int [ds] e^{ 2 \pi i (s_1 + s_3 - s_2 - s_4 )} \, = \, - {{ r^4 } \over {16}} (1 + {{3 i} \over {\pi }} ).
\end{eqnarray} 
Summing ${\cal I}_1$ and ${\cal I}_2$, we finally get
\begin{equation}
\label{fin}
\lim_{\theta \to \infty} \WW_4^{(cr)} \, = \, - {{{ A}^2 }\over {8 {\pi }^2 }}. 
\end{equation}

The conclusion is that a perturbative Euclidean calculation with the WML
prescription is feasible and leads to a regular result, in spite of the
fact that non-commutativity in the Minkowski formulation involves the time variable. 
Abelian-like exponentiation does not hold, not surprisingly for $N>1$,
as the same phenomenon occurs also in the ordinary commutative
case (see Eq.(\ref{wml}))\cite{bbg,anlu}. 
However one should remark that the presence of the Moyal
phase introduces a $\theta$-dependent term in $\WW_4$ that changes 
sign from the two limiting values $\theta=0$ and $\theta \to \infty$.
A destructive interference is likely to occur.

It would be very interesting in our opinion to find an algorithm for
computing higher perturbative orders, generalizing to the non-commutative
theory the matrix model 
occurring in the ordinary case \cite{basvil}. However it would require a full 
control of the Moyal phases for diagrams of increasing complexities.
Alternatively one could try to envisage an algorithm of a non-perturbative
type. This is likely to be possible starting from a suitable brane 
formulation and will be explored in future investigations.

\section{The Minkowski formulation}

\noindent

In this section we face the problem of performing a perturbative ${\cal O}(g^4)$
calculation of the Wilson loop, without making use of an Euclidean
formulation. This is necessary as  eventually we want to consider 't Hooft's 
prescription for the vector propagator, which cannot be Wick-rotated.

We start considering again the WML case, but refrain from
passing to Euclidean coordinates.
The ${\cal O}(g^4)$ contribution from the crossed diagram reads
\begin{eqnarray}
\label{mink}
\WW_{4}^{(cr)}\ &=&- \ \int [ds] \dot{x}_{-}(s_1)\ldots\dot{x}_{-}(s_{4})
\int \frac{dp_{+} dp_{-}}{4\pi^2 [p_{-}^2]} \frac{dq_{+} dq_{-}}{4\pi^2 
[q_{-}^2]}
e^{i\theta (p_{-}q_{+}-q_{-}p_{+})} \nonumber \\
&\times& \exp(i[p_{+}(x_{-}(s_1)-x_{-}(s_3))+p_{-}(x_{+}(s_1)-x_{+}(s_3))])\nonumber \\
&\times& \exp(i[q_{+}(x_{-}(s_2)-x_{-}(s_4))+q_{-}(x_{+}(s_2)-x_{+}(s_4))]),
\end{eqnarray}
where the denominators $[p_{-}^{-2}]$ and $[q_{-}^{-2}]$ 
have to be interpreted according to Eq.(\ref{mand}).

Following the theory of distributions, the equation above can be rewritten as
\begin{eqnarray}
\label{mink1}
\WW_{4}^{(cr)}\ &=& \  \int [ds] \dot{x}_{-}(s_1)\ldots\dot{x}_{-}(s_{4})
\int \frac{dp_{+} dp_{-}}{4\pi^2 [p_{-}]} \frac{dq_{+} dq_{-}}{4\pi^2 
[q_{-}]}
e^{i\theta (p_{-}q_{+}-q_{-}p_{+})} \nonumber \\
&\times& \exp(i[p_{+}(x_{-}(s_1)-x_{-}(s_3))+p_{-}(x_{+}(s_1)-x_{+}(s_3))])\nonumber \\
&\times& \exp(i[q_{+}(x_{-}(s_2)-x_{-}(s_4))+q_{-}(x_{+}(s_2)-x_{+}(s_4))])\nonumber \\
&\times&\Big( (x_{+}(s_1)-x_{+}(s_3)+\theta \ q_{+}\Big)\Big( (x_{+}(s_2)-x_{+}(s_4)
-\theta \ p_{+}\Big).
\end{eqnarray}

We now recall the well-known relation between the WML distribution and 
the Cauchy principal value 
\beq
\label{split}
[k_{-}+i\epsilon k_{+}]^{-1}= 
[k_{-}^{-1}]_{PV}-i\pi\ \delta (k_{-})\ sign(k_{+}).
\eeq

Eq.(\ref{mink1}) splits naturally into three contributions, the first one being 
nothing but 
$\WW_4^{(cr)}$ evaluated with 't Hooft's prescription
\begin{eqnarray}
\label{pv1}
\WW_{4}^{(cr,1)}\ &=& \  \int [ds] \dot{x}_{-}(s_1)\ldots\dot{x}_{-}(s_{4})
\int \frac{dp_{+} dp_{-}}{4\pi^2 [p_{-}]_{PV}} \frac{dq_{+} dq_{-}}{4\pi^2 
[q_{-}]_{PV}}
e^{i\theta (p_{-}q_{+}-q_{-}p_{+})} \nonumber \\
&\times& \exp(i[p_{+}(x_{-}(s_1)-x_{-}(s_3))+p_{-}(x_{+}(s_1)-x_{+}(s_3))])\nonumber \\
&\times& \exp(i[q_{+}(x_{-}(s_2)-x_{-}(s_4))+q_{-}(x_{+}(s_2)-x_{+}(s_4))])\nonumber \\
&\times&\Big( (x_{+}(s_1)-x_{+}(s_3)+\theta \ q_{+}\Big)\Big( (x_{+}(s_2)-x_{+}(s_4)
-\theta \ p_{+}\Big).
\end{eqnarray}
It is easy to realize that the expression above vanishes at $\theta=0$ owing
to the measure $[ds]$. When $\theta \ne 0$
the integrations over the momenta can be performed, leading to the formal
result
\begin{eqnarray}
\label{formal}
&&\WW_{4}^{(cr,1)}\ =- \frac{\theta^2}{4\pi^2} \ \int [ds] 
\frac{\dot{x}_{-}(s_1)\ldots\dot{x}_{-}(s_{4})}{[\Big(x_{-}(s_1)-x_{-}(s_3)\Big)^2]_{PV}
[\Big(x_{-}(s_2)-x_{-}(s_4)\Big)^2]_{PV}} \\
&\times& \exp \frac{i}{\theta}\Big[\Big(x_{-}(s_1)-x_{-}(s_3)\Big)\Big(x_{+}(s_2)-x_{+}
(s_4)\Big)-
\Big(x_{+}(s_1)-x_{+}(s_3)\Big)\Big(x_{-}(s_2)-x_{-}(s_4)\Big)\Big],\nonumber
\end{eqnarray}
where the denominators have to be interpreted as derivatives of the 
Cauchy principal value distribution.

The second contribution comes from the product of the two
$\delta$-distributions
\begin{eqnarray}
\label{delta2}
&&\WW_{4}^{(cr,2)}\ =- \frac{1}{16\pi^2}\  \int [ds] 
\dot{x}_{-}(s_1)\ldots\dot{x}_{-}(s_{4})
\int dp_{+} dq_{+}\ sign(p_{+})\ sign(q_{+})
\nonumber \\
&\times& \exp(i[p_{+}(x_{-}(s_1)-x_{-}(s_3))])\, \exp(i[q_{+}(x_{-}(s_2)-x_{-}(s_4))]
\nonumber \\
&\times&\Big( (x_{+}(s_1)-x_{+}(s_3)+\theta \ q_{+}\Big)\Big( (x_{+}(s_2)-x_{+}(s_4)
-\theta \ p_{+}\Big)= \nonumber \\
&=& \frac{1}{4\pi^2}\  \int [ds] 
\dot{x}_{-}(s_1)\ldots\dot{x}_{-}(s_{4})\Big[\frac{\Big(x_{+}(s_1)-x_{+}(s_3)\Big)
\Big(x_{+}(s_2)-x_{+}(s_4)\Big)}{\Big(x_{-}(s_1)-x_{-}(s_3)\Big)_{PV}
\Big(x_{-}(s_2)-x_{-}(s_4)\Big)_{PV}} \nonumber \\
&-&i\theta \frac{x_{+}(s_1)-x_{+}(s_3)}{\Big[ \Big(x_{-}(s_1)-x_{-}(s_3)\Big)^2
\Big]_{PV} \Big[ \Big(x_{-}(s_2)-x_{-}(s_4)\Big)\Big]_{PV}}\nonumber \\
&+&i\theta \frac{x_{+}(s_2)-x_{+}(s_4)}{\Big[ \Big(x_{-}(s_2)-x_{-}(s_4)\Big)^2
\Big]_{PV} \Big[ \Big(x_{-}(s_1)-x_{-}(s_3)\Big)\Big]_{PV}} \nonumber \\ 
&+&\theta^2 \frac{1}{\Big[ \Big(x_{-}(s_1)-x_{-}(s_3)\Big)^2
\Big]_{PV} \Big[ \Big(x_{-}(s_2)-x_{-}(s_4)\Big)^2 \Big]_{PV}} \ \Big].
\end{eqnarray}

Finally there is the term due to mixed products (see Eq.(\ref{split}))
\begin{eqnarray}
\label{mixed}
\WW_{4}^{(cr,3)}\ &=&- i\pi \int [ds] \dot{x}_{-}(s_1)\ldots\dot{x}_{-}(s_{4})
\int \frac{dp_{+} dp_{-}}{4\pi^2} \frac{dq_{+} dq_{-}}{4\pi^2}
e^{i[\theta (p_{-}q_{+}-q_{-}p_{+})]} \nonumber \\
&\times& \Big[ \frac{sign(p_{+})\ \delta(p_{-})}{[q_{-}]_{PV}}+
\frac{sign(q_{+})\ \delta(q_{-})}{[p_{-}]_{PV}}\Big] \nonumber \\
&\times& \exp(i[p_{+}(x_{-}(s_1)-x_{-}(s_3))+p_{-}(x_{+}(s_1)-x_{+}(s_3))])\nonumber \\
&\times& \exp(i[q_{+}(x_{-}(s_2)-x_{-}(s_4))+q_{-}(x_{+}(s_2)-x_{+}(s_4))])\nonumber \\
&\times&\Big( (x_{+}(s_1)-x_{+}(s_3)+\theta \ q_{+}\Big)\Big( (x_{+}(s_2)-x_{+}(s_4)
-\theta \ p_{+}\Big).
\end{eqnarray}
This term vanishes for symmetry reasons.

Now the (singular) $\theta$-dependent terms in $\WW_{4}^{(cr,2)}$ cancel
against the analogous (singular) terms which would occur when expanding
the exponential in $\WW_{4}^{(cr,1)}$, namely

\begin{equation}
\label{primo}
\WW_{4}^{(cr,1)}|_{I}\ =-\frac{\theta^2}{4\pi^2} \int
[ds]\  \frac{\dot{x}_{-}(s_1)\ldots\dot{x}_{-}(s_{4})}
{\Big[ \Big(x_{-}(s_1)-x_{-}(s_3)\Big)^2
\Big]_{PV} \Big[ \Big(x_{-}(s_2)-x_{-}(s_4)\Big)^2 \Big]_{PV}} 
\end{equation}
and
\begin{eqnarray}
\label{secondo}
\WW_{4}^{(cr,1)}|_{II}\ &=& \frac{-i\theta}{4\pi^2} 
\int [ds]\ \dot{x}_{-}(s_1)\ldots\dot{x}_{-}(s_{4})
\Big( \frac{x_{+}(s_2)-x_{+}(s_4)}{\Big[ \Big(x_{-}(s_2)-x_{-}(s_4)\Big)^2
\Big]_{PV} \Big[ \Big(x_{-}(s_1)-x_{-}(s_3)\Big)\Big]_{PV}}\nonumber \\
&-& \frac{x_{+}(s_1)-x_{+}(s_3)}{\Big[ \Big(x_{-}(s_1)-x_{-}(s_3)\Big)^2
\Big]_{PV} \Big[ \Big(x_{-}(s_2)-x_{-}(s_4)\Big)\Big]_{PV}}\Big). 
\end{eqnarray}

In this way $\WW_4^{(cr)}$ turns out to be 
regular, as expected. 

At $\theta=0$ it is represented by the first term of
Eq.(\ref{delta2}): the quantity 
\begin{equation}
\label{deltar}
\WW_{4}^{(cr)}(\theta=0)\ = 
\  \frac{1}{4\pi^2}\  \int [ds]\ 
\dot{x}_{-}(s_1)\ldots\dot{x}_{-}(s_{4})\ \frac{\Big(x_{+}(s_1)-x_{+}(s_3)\Big)
\Big(x_{+}(s_2)-x_{+}(s_4)\Big)}{\Big(x_{-}(s_1)-x_{-}(s_3)\Big)
\Big(x_{-}(s_2)-x_{-}(s_4)\Big)} 
\end{equation}
is nothing but the usual
WML contribution as it should, after 
the Minkowski replacement ${\cal A}\to i{ A}.$ From here on it is no longer necessary
to prescribe denominators as the measure provides the necessary regularization.

In order to compute the large-$\theta$ limit of Eq.(\ref{mink1}), we remember that
the first two terms if we expand the exponential in Eq.(\ref{formal}), namely
Eqs.(\ref{primo},\ref{secondo}), cancel
against the corresponding ones in Eq.(\ref{delta2}). The next ($\theta$-independent)
term in such an expansion
\begin{eqnarray}
\label{formalg}
&&\WW_{4}^{(cr,1)}|_{III}\ =\ \  \frac{1}{8\pi^2} \ \int [ds]\ 
\frac{\dot{x}_{-}(s_1)\ldots\dot{x}_{-}(s_{4})}{[\Big(x_{-}(s_1)-x_{-}(s_3)\Big)^2]
[\Big(x_{-}(s_2)-x_{-}(s_4)\Big)^2]} \\
&\times& \Big[\Big(x_{-}(s_1)-x_{-}(s_3)\Big)\Big(x_{+}(s_2)-x_{+}
(s_4)\Big)-
\Big(x_{+}(s_1)-x_{+}(s_3)\Big)\Big(x_{-}(s_2)-x_{-}(s_4)\Big)\Big]^2 \nonumber
\end{eqnarray}
has to be added to the surviving one in Eq.(\ref{delta2}), to obtain
\begin{eqnarray}
\label{formalgg}
&&\WW_{4}^{(cr)}|_{\theta \to \infty}\ =\ \  \frac{1}{8\pi^2} \ \int [ds]\ 
\frac{\dot{x}_{-}(s_1)\ldots\dot{x}_{-}(s_{4})}{[\Big(x_{-}(s_1)-x_{-}(s_3)\Big)^2]
[\Big(x_{-}(s_2)-x_{-}(s_4)\Big)^2]} \\
&\times& \Big(\Big[\Big(x_{-}(s_1)-x_{-}(s_3)\Big)\Big(x_{+}(s_2)-x_{+}
(s_4)\Big)\Big]^2+
\Big[\Big(x_{+}(s_1)-x_{+}(s_3)\Big)\Big(x_{-}(s_2)-x_{-}(s_4)\Big)\Big]^2\Big),\nonumber
\end{eqnarray}
all other terms in the expansion of Eq.(\ref{formal}) vanishing in the large-$\theta$
limit.

The integrals in Eq.(\ref{formalgg}) can be fairly easily computed, 
using for instance a rectangular contour; the final outcome is
\begin{equation}
\label{finres}
\WW_{4}^{(cr,1)}|_{III}\ =\ {{{\cal A}^2 }\over {8 {\pi }^2 }},
\end{equation}
the opposite sign with respect to the one in Eq.(\ref{fin}) being due to
the Minkowski replacement ${\cal A}\to i{ A}.$

We stress that here the calculation has been performed directly in a
Minkowski context, without rotating the Moyal phase.

The quantities $\WW_{4}^{(cr,1)}|_{I}$ and $\WW_{4}^{(cr,1)}|_{II}$ occurring in the 
Wilson loop calculation with 't Hooft's prescription, are
genuinely singular: integration of the distributions  in 
Eqs.(\ref{primo},\ref{secondo}) cannot be
defined with the measure $[ds]\ \dot{x}_{-}(s_1)\ldots\dot{x}_{-}(s_{4})$. If
a cutoff is introduced as a regulator, the expressions diverge unavoidably
when the cutoff is removed. In particular Eq.(\ref{primo}) does not depend on the area.

Those quantities are therefore responsible of the failure of
the Wilson loop calculation in Minkowski space if 't Hooft's prescription
is adopted for the propagator. However we find quite remarkable that
singularities are precisely confined within these two terms. The rest
of the expression is regular in spite of the singular nature of 't
Hooft's propagator and coincides with the result one obtains
after analytic continuation starting
from the $WML$ propagator. 
We find this unexpected connection very intriguing when remembering
that in ordinary theories the difference between the two formulations on a 
compact manifold
concerns their topological structure: the $WML$ formulation only
captures the topologically trivial sector \cite{anlu}.

\section{Conclusions}

\noindent
In usual commutative two-dimensional Yang-Mills theories ($YM_{1+1}$), closed
Wilson loops can be exactly computed, either using geometrical techniques
or by summing a perturbative series in light-cone gauge, in which the
propagator is handled as a Cauchy principal value prescription.
As a matter of fact in this case contributions from crossed diagrams vanish
thanks to peculiar support conditions. Only planar diagrams survive, enforcing 
the picture of the loop as an exchange potential between two (static)
colour sources. A simple area exponentiation is the final outcome.

If the propagator is instead prescribed in a causal way, crossed diagrams no 
longer vanish; the result one obtains by summing the series no longer 
exhibits Abelian-like exponentiation for $N>1$ and only captures the zero instanton
sector of the theory \cite{anlu}.

When considering non-commutative $YM_{1+1}$, one may wonder whether such a theory 
should make sense at all, owing to the pathologies occurring when non-commutativity
involves the time variable. Indeed even simple models exhibit either trivial
or inconsistent solutions in such a case \cite{seme}.

Therefore we find quite remarkable that a sensible expression can be obtained
for a closed Wilson loop as a fourth-order
perturbative calculation using the $WML$ propagator. 
It exhibits a smooth limit when the non-commutative
parameter $\theta$ tends to zero, thereby recovering the usual commutative result;
in the large non-commutative limit $\theta \to \infty$, the crossed diagrams
contribution changes sign,
thus suggesting the possibility of a critical value of $\theta$.  
The loop does not obey the Abelian-like
exponentiation constraint, but this is not surprising for $N>1$, as it happens also
in the usual commutative case. For $N=1$ the commutative Wilson loop trivially
exponentiates; this does not happen in the noncommutative theory, providing
a further example of the deep difference between the two cases.

Of course it would be very interesting in our opinion to be able to generalize such
a calculation beyond the ${\cal O}(g^4)$-order and to sum the resulting
perturbative series as it was done in ref.\cite{stau} in the ordinary case.
Here the difficulty lies in the fact that different patterns of crossing entail
different Moyal phases; no combinatorial algorithm has been found so far to 
cope with them, to our knowledge. It remains an open problem.

More dramatic is the situation when considering  't Hooft's form of
the free propagator, which does not allow a smooth transition to Euclidean
variables. Here the presence of the Moyal phase produces singularities
even at ${\cal O}(g^4)$, which cannot be cured. Strictly speaking, the
Wilson loop diverges (at least perturbatively). Nevertheless we find
extremely interesting the circumstance that the difference between its Moyal
dependent term and the analogous one in the $WML$ case turns out to be 
confined only in the two
singular terms $\WW_{4}^{(cr,1)}|_{I}$ and $\WW_{4}^{(cr,1)}|_{II}$, 
which in turn possess a trivial $\theta$-dependence. This
might entail far-reaching consequences
on the topological structure of the theory, when considered on a compact
manifold.

These features are currently under investigation
and the results will be reported elsewhere.

$\bf Acknowledgements$

We thank L. Griguolo and F. Vian for discussions at an early stage
of this work.


\begin{thebibliography}{99}
\bibitem{cds} A.~Connes, M.R.~Douglas and A.~Schwarz, JHEP \ {\bf 02} 
(1998) 003.
\bibitem{dh} M.R.~Douglas and C.~Hull, JHEP \ {\bf 02}  (1998) 008.
\bibitem{seiwi} N.~Seiberg and E.~Witten, JHEP\ {\bf 09}  (1999) 032.
\bibitem{minwa}S.~Minwalla, M.~Van Raamsdonk and N.~Seiberg, JHEP \
{\bf 02} (2000) 020.
\bibitem{suski}A.~Matusis, L.~Susskind and N.~Toumbas, JHEP \ {\bf 12} 
(2000) 002.
\bibitem{gomis} J.~Gomis and T.~Mehen, Nucl. \ Phys.\ {\bf B591} 
(2000) 265.
\bibitem{stringhi}
O.~Andreev and H.~Dorn,
Nucl.\ Phys.\ {\bf B583} (2000) 145; 
Y.~Kiem and S.~Lee,
Nucl.\ Phys.\ {\bf B586} (2000) 303;
A.~Bilal, Chong-Sung~Chu and R.~Russo,
Nucl.\ Phys.\ {\bf B582} (2000) 65; 
J.~Gomis, M.~Kleban, T.~Mehen, M.~Rangamani and S.~Shenker,
JHEP \ {\bf 08} (2000) 011.
\bibitem{kamimura}
J.~Gomis, K.~Kamimura and J.~Llosa,
Phys.\ Rev.\  {\bf D63} (2001) 045003;
J.~Gomis, K.~Kamimura and T.~Mateos,
JHEP \ {\bf 03} (2001) 010.
\bibitem{om}N.~Seiberg, L.~Susskind and N.~Toumbas, JHEP \ {\bf 06} 
(2000) 021;
R.~Gopakumar, J.~Maldacena, S.~Minwalla and A.~Strominger, JHEP \ {\bf
06}  (2000) 036.
\bibitem{barb}J.~L.~Barbon and E.~Rabinovici,
Phys.\ Lett.\ {\bf B486} (2000) 202.
\bibitem{alva}L.~Alvarez-Gaume, J.L.F.~Barbon and R.~Zwicky,
{\it ``Remarks on Time-Space Noncommutative Field Theories''},
{\tt hep-th/0103069}.
\bibitem{bgnv}A.~Bassetto, L.~Griguolo, G.~Nardelli and F.~Vian,
{\it ``On the unitarity of quantum gauge theories on non-commutative spaces''},
{\tt hep-th/0105257}.
\bibitem{test}A.~Bassetto, G.~Nardelli and R.~Soldati, {\it ``Yang-Mills Theories
in Algebraic Non-covariant Gauges''}, World Scientific, Singapore 1991.
\bibitem{gross}D.~J.~Gross, A.~Hashimoto and N.~Itzhaki,
{\it ``Observables of Non-Commutative Gauge Theories''}, {\tt hep-th/0008075}.
\bibitem{alvar}L.~Alvarez-Gaume and S.~R.~Wadia,
Phys.\ Lett.\ {\bf B501} (2001) 319.
\bibitem{dorn}M.~Abou-Zeid and H.~Dorn,
Phys.\ Lett.\ {\bf B504} (2001) 165.
\bibitem{ambj}J.~Ambjorn, Y.~M.~Makeenko, J.~Nishimura and R.~J.~Szabo,
Phys.\ Lett.\ {\bf B480} (2000) 399; JHEP \ {\bf 05} (2000) 023.
\bibitem{stau}M.~Staudacher and W.~Krauth, Phys.\ Rev.\ {\bf D57} (1998) 2456.
\bibitem{hoo}G.~'t Hooft, Nucl.\ Phys.\ {\bf B75} (1974) 461.
\bibitem{wum}T.T.~Wu, Phys.\ Lett.\ {\bf 71B} (1977) 142; S.~Mandelstam, Nucl.\
Phys.\ {\bf B213} (1983) 149; G.~Leibbrandt, Phys.\ Rev.\ {\bf D29} (1984) 1699.
\bibitem{bbg}A.~Bassetto, F.~De Biasio and L.~Griguolo, Phys. Rev. Lett. {\bf 72}
(1994) 3141.
\bibitem{anlu}A.~Bassetto and L.~Griguolo, Phys. Lett. {\bf B443} (1998) 325.
\bibitem{sheikh}A.~Das and M.M.~Sheikh-Jabbari, {\it ``Absence of Higher Order 
Corrections to Noncommutative Chern-Simons Coupling''}, hep-th/0103139.
\bibitem{basvil}A.~Bassetto, Nucl. Phys. {\bf B} (Proc. Suppl.) {\bf 88} (2000) 184.
\bibitem{seme}E.T.~Akhmedov, P.~DeBoer and G.W.~Semenoff, JHEP {\bf 0106} (2001) 009.




\end{thebibliography}
\end{document}